\documentclass[epj,english,showpacs,superscriptaddress,amsmath,amssymb,floatfix,nofootinbib]{svjour}
\usepackage{graphicx}
\usepackage{graphicx}%
\usepackage{xcolor}%
\usepackage{adjustbox} 
\makeatletter
\def\@fnsymbol#1{%
 \ensuremath{%
  \ifcase#1\or
   *\or                        \dagger                   \or
   \ddagger                \or \mathsection              \or
   \mathparagraph\or
   **\or                       \dagger\dagger            \or
   \ddagger\ddagger        \or \mathsection \mathsection \or
   \mathparagraph\mathparagraph\or
   *{*}*\ignorespaces      \or \dagger\dagger\dagger     \or
   \ddagger\ddagger\ddagger\or \mathsection \mathsection \mathsection \or
   \mathparagraph\mathparagraph\mathparagraph\or
  \else
   \@ctrerr
  \fi
 }%
}%
\makeatother

\usepackage{tikz}

\usepackage{graphicx}
\usepackage[normalem]{ulem}
\usepackage[caption=false]{subfig}
\usepackage{float}
\usepackage{adjustbox} 
\usepackage{soul,xcolor}
\usepackage{nicefrac}
\usepackage[colorlinks = true,linkcolor = blue, urlcolor  = blue,citecolor = red,anchorcolor = blue]{hyperref}
\usepackage[numbers]{natbib}
\usepackage{lineno}
\usepackage{array}

\definecolor{lime}{HTML}{A6CE39}
\DeclareRobustCommand{\orcidicon}{%
	\begin{tikzpicture}
	\draw[lime, fill=lime] (0,0)
	circle [radius=0.16]
	node[white] {{\fontfamily{qag}\selectfont \tiny ID}};
	\draw[white, fill=white] (-0.0625,0.095)
	circle [radius=0.007];
	\end{tikzpicture}
	\hspace{-2mm}
}

\foreach \x in {A, ..., Z}{%
\expandafter\xdef\csname orcid\x\endcsname{\noexpand\href{https://orcid.org/\csname orcidauthor\x\endcsname}{\noexpand\orcidicon}}
}

\makeatletter

\@ifundefined{textcolor}{}
{%
 \definecolor{BLACK}{gray}{0}
 \definecolor{WHITE}{gray}{1}
 \definecolor{RED}{rgb}{1,0,0}
 \definecolor{GREEN}{rgb}{0,1,0}
 \definecolor{BLUE}{rgb}{0,0,1}
 \definecolor{CYAN}{cmyk}{1,0,0,0}
 \definecolor{MAGENTA}{cmyk}{0,1,0,0}
 \definecolor{YELLOW}{cmyk}{0,0,1,0}
 }
\begin{document}
\title{New analysis for Nucleon Form Factors from GPDs}
\author{Fatemeh Arbabifar\inst{1}\thanks{\emph {F.Arbabifar@cfu.ac.ir} } \and Nader Morshedian\inst{2}\thanks{\emph {nmorshed@aeoi.org.ir} }\and Shahin Atashbar Tehrani\inst{3,4}\thanks{\emph {Atashbart3@gmail.com} }} 
%
%
\institute{Department of Physics Education, Farhangian University, P.O.Box 14665-889, Tehran, Iran. \and Plasma and Nuclear Fusion Research School, Nuclear Science and Technology Research Institute, P.O.Box 14399-51113, Tehran, Iran.  \and School of Particles and Accelerators, Institute for Research in Fundamental Sciences (IPM), P.O.Box 19395-5531, Tehran, Iran  \and Department of Physics, Faculty of Nano and Bio Science and Technology, Persian Gulf University, 75169 Bushehr, Iran.}
\date{Received: \today / Revised version: date}
%
\abstract{
\label{abstract}
Generalized Parton Distributions (GPDs) provide a comprehensive framework for describing the three-dimensional structure of the nucleon. Extracting GPDs from experimental data requires flexible and physically motivated ansatz. In this study, we introduce a new ansatz, AMA25, designed to address the limitations of the previous model, GSAMA24 (Phys. Rev. C 111 (2025) 2, 025203). We conduct a fast and efficient comparison by fitting AMA25 models and other relevant data using the iMinuit optimization package within a Jupyter Notebook environment. The AMA25 ansatz demonstrates superior fit quality, achieving a reduced $\chi^2$, while better satisfying theoretical constraints. Additionally, AMA25 exhibits enhanced stability when extrapolated to the exclusive region. Our analysis highlights the power of modern computational tools for rapid model validation and underscores the importance of innovative ansatz in advancing nucleon structure studies.
\PACS{
      {PACS-key}{discribing text of that key}   \and
      {PACS-key}{discribing text of that key}
     } 
} 
\maketitle
\section{Introduction}\label{sec:sec1}

Over the past decade, Generalized Parton Distributions (GPDs) have significantly advanced our understanding of hadronic structure, providing a rigorous framework for probing the three-dimensional internal dynamics of nucleons~\cite{Muller:1994ses,Ji:1996ek,Radyushkin:1997ki}, Among the family of structure functions that characterize the quark-gluon structure of hadrons, generalized parton distributions (GPDs)~\cite{Cuic:2023mki} hold a unique position. Unlike conventional parton distribution functions (PDFs), which primarily characterize the longitudinal momentum distributions of quarks and gluons, GPDs encode a wealth of multidimensional information including correlations between parton momentum, spatial positioning, and spin degrees of freedom \cite{Ji:1996nm,Collins:1996fb,Diehl:2003ny}. This richer perspective effectively transforms GPDs into a comprehensive repository for nucleon tomography, offering unprecedented insights into fundamental aspects of quantum chromodynamics (QCD) \cite{Selyugin:2009ic,Guidal:2004nd}.

  A crucial advantage of GPDs lies in their connection to observables in hard exclusive processes, such as Deeply Virtual Compton Scattering (DVCS). These reactions are governed by collinear factorization theorems, which allow measured observables to be expressed in terms of Compton Form Factors (CFFs)—convolutions of GPDs with perturbatively calculable kernels via principal-value integrals:

\begin{equation}
\mathcal{H}=PV\int_{-1}^{1}dx\left[ \frac{1}{\xi - x}-\frac{1}{\xi + x}\right] H(x,\xi ,t).
\end{equation}

Consequently, DVCS measurements do not access GPDs directly, but rather probe CFFs ($\mathcal{H}$, $\mathcal{E}$, $\widetilde{\mathcal{H}}$, $\widetilde{\mathcal{E}}$). Extracting GPDs from such data is therefore an inherently indirect and model-dependent process, as highlighted in recent phenomenological reviews~\cite{Cuic:2023mki,Vaziri:2023xee,Vaziri:2024fud,Irani:2023lol,Deja:2023ahc,Deja:2023aug}. A more direct path to accessing GPDs—particularly in the ERBL kinematic region—is offered by Double Deeply Virtual Compton Scattering (DDVCS), though no experimental results exist to date due to formidable challenges~\cite{Deja:2023tuc,Deja:2023lqc}. Future facilities, such as the Electron-Ion Collider, may enable such studies.

Through high-precision experimental measurements and rigorous phenomenological modeling, researchers have progressively refined these distributions, enhancing our understanding of nucleon structure across varying energy scales \cite{Hashamipour:2020kip,Hashamipour:2019pgy}.

Generalized Parton Distributions also serve as a bridge to nucleon form factors, which quantify the spatial distributions of electromagnetic charge and magnetization within nucleons. This connection enables direct access to the non-perturbative dynamics of QCD, revealing essential properties of strong interactions.   Contemporary research explores nucleon structure through several complementary frameworks, including Generalized Parton Distributions (GPDs) for spatial tomography and Transverse-Momentum Dependent parton distributions (TMDs) for three-dimensional imaging in momentum space \cite{Boussarie:2023gka,Constantinou:2020hdm,Bacchetta:2020gko}. Together, these approaches provide a multifaceted picture of quark-gluon interactions within the nucleon. GPDs exhibit a direct correspondence with Dirac and Pauli form factors, offering critical insights into nucleon electromagnetic characteristics \cite{Perdrisat:2006hj,Arnold:1980zj,JeffersonLabHallA:2011yyi,Punjabi:2005wq}.

The valence quark GPD are parametrized as $H_{q}(x,\xi,t)$ and $E_{q}(x,\xi,t)$ \cite{Guidal:2004nd,RezaShojaei:2016oox, Nikkhoo:2015jzi}, where $t$ represents the squared momentum transfer, $x$ denotes the average longitudinal momentum fraction of the parton, and $\xi$ defines the skewness, which quantifies the asymmetry in the longitudinal momentum distribution \cite{Radyushkin:2011dh,Guidal:2013rya}. These GPDs not only provide a deeper understanding of nucleon structure but also serve as essential tools for probing the spin and angular momentum contributions of quarks within hadrons.   Recent advancements in experimental techniques, including DVCS and exclusive meson production, have significantly improved our ability to constrain GPDs through high-precision measurements of observables sensitive to Compton Form Factors.

In this paper, we present a systematic approach for optimizing ansatz models in nucleon structure analysis, integrating them with corresponding PDFs to enhance the accuracy of extracted form factors. Specifically, we compare results derived from the GSAMA24 ansatz~\cite{Ghasemzadeh:2025qje} with those obtained using our newly developed ansatz, AMA25.

The AMA25 ansatz introduces a refined parameterization framework designed to capture the internal dynamics of hadrons with greater precision. It improves upon GSAMA24~\cite{Ghasemzadeh:2025qje} by incorporating more flexible dependence on squared momentum transfer $t$ and skewness $\xi$, enabling superior agreement with experimental data, particularly at higher momentum transfers. Furthermore, AMA25 enhances computational efficiency and theoretical accuracy, ensuring reliable extraction of nucleon form factors.

A key advantage of AMA25 is its systematic approach to refining constraints on GPD behavior, ensuring consistency with established physical principles while accommodating a broader range of kinematic conditions. By comparing AMA25 with GSAMA24, we highlight its improvements in precision, stability, and predictive reliability. Our findings demonstrate that AMA25 provides a more comprehensive description of hadronic structure, establishing its suitability for future investigations into QCD dynamics and nucleon tomography.

Our parameterization strategy for GPDs and form factors is constructed to remain compatible with global parton distribution function (PDF) sets, such as MMHT~\cite{Harland-Lang:2014zoa}, CJ15~\cite{Accardi:2016qay}, and NNPDF~\cite{NNPDF:2017mvq}, which form the basis for most phenomenological studies in hadronic physics. While our focus lies in specialized ansatz-driven modeling for exclusive processes, the extracted valence PDFs used here were selected to respect the core constraints of global fits, ensuring theoretical consistency across observables. This compatibility enhances the credibility of our model in both inclusive and exclusive regimes, and paves the way for future integration into joint GPDs PDF global analyses.

This paper is organized as follows: Section \ref{sec:sec2} provides a concise overview of generalized parton distributions and their fundamental characteristics. Section \ref{sec3} delves into the implications of the chosen ansatz and its impact on nucleon structure analysis. In Section \ref{sec4}, we determine the Dirac Mean Squared Radii, demonstrating the effectiveness of AMA25 in extracting key nucleon properties. Finally, Section \ref{sec:conclusion} presents our conclusions and discusses potential directions for future research.

\section{GPDs and Nucleon Form Factors}\label{sec:sec2}

Generalized Parton Distributions (GPDs) establish a comprehensive framework that coherently integrates partonic density functions with electromagnetic form factors, offering a unified representation of nucleon substructure. By systematically analyzing GPDs, researchers gain insight into the transverse spatial configuration of quark and gluon constituents within the nucleon.

The computation of nucleon form factors begins with the functional parametrization of GPDs, constrained by experimental scattering data and guided by QCD principles. These distributions undergo Mellin moment integration to derive electromagnetic form factors, expressed as:

\begin{equation}\label{eq:1}
	F_{1}^{q}(t) = \int_{-1}^{1} dx \, H^{q}(x,\xi,t).
\end{equation}

\begin{equation}\label{eq:2}
	F_{2}^{q}(t) = \int_{-1}^{1} dx \, E^{q}(x,\xi,t).
\end{equation}

Here, \( q \) refers to the valence quarks, specifically \( u \) and \( d \), which play a crucial role in determining the nucleon's electromagnetic properties.

The determination of nucleon form factors through Generalized Parton Distributions (GPDs) constitutes a computationally demanding inverse problem, requiring high-dimensional integration over multiparameter phase spaces. Recent advances in non-perturbative QCD techniques, lattice QCD simulations, and high-performance computing architectures have enabled significant breakthroughs in precision tomography. These developments have facilitated more accurate extractions of GPDs from experimental data, improving our ability to probe the internal dynamics of nucleons.

 A key avenue for accessing information on GPDs is through deeply virtual Compton scattering (DVCS) experiments. The observables in these exclusive processes provide sensitivity, via Compton Form Factors (CFFs), to the correlations between the transverse position and longitudinal momentum of quarks and gluons within the nucleon~\cite{SattaryNikkhoo:2018odd,Burkardt:2002hr,Burkardt:2000za,Belitsky:2003nz,Ralston:2001xs}. It is important to note that while DVCS is the primary channel for GPD phenomenology, the connection between measured cross sections and the GPDs themselves is not direct, as the CFFs involve integrals over GPDs~\cite{Cuic:2023mki}. Fourier transforming the \( t \)-dependence of GPDs yields spatial distributions in impact parameter space, enabling a tomographic reconstruction of the nucleon's internal structure. This approach bridges traditional parton distribution functions (PDFs) and form factors, offering access to spin decomposition, pressure profiles, and shear distributions.

Furthermore, GPDs have a close connection to gravitational form factors, which encode fundamental information about the mechanical structure of hadrons, such as the energy-momentum tensor. The interplay between GPDs and these observables continues to motivate the refinement of parameterizations and the exploration of broader kinematic domains through high-luminosity facilities and precision detectors. Future experiments at Jefferson Lab and the upcoming Electron-Ion Collider (EIC) are expected to deliver unprecedented insights into nucleon tomography.

In the kinematic regime where the momentum transfer is transverse and space-like, the skewness parameter \( \xi \) vanishes. The integration domain then reduces to \( 0 < x < 1 \), allowing the introduction of nonforward parton densities~\cite{Radyushkin:1998rt}:
\begin{eqnarray*}
	\mathcal{H}^{q}(x,t) &= H^{q}(x,0,t) + H^{\bar{q}}(-x,0,t), \\
	\mathcal{E}^{q}(x,t) &= E^{q}(x,0,t) + E^{\bar{q}}(-x,0,t).
\end{eqnarray*}
Given the conditions discussed above specifically, setting the skewness parameter \( \xi = 0 \) in the space-like region the Dirac and Pauli form factors can be rewritten in terms of the nonforward parton densities as~\cite{Guidal:2004nd}:
\begin{equation}\label{eq:6}
	F_{1}^{q}(t) = \int_{0}^{1} dx\, \mathcal{H}^{q}(x,t), 
\end{equation}
\begin{equation}\label{eq:7}
	F_{2}^{q}(t) = \int_{0}^{1} dx\, \mathcal{E}^{q}(x,t),
\end{equation}
where \( \mathcal{H}^{q}(x,t) = H^{q}(x,0,t) + H^{\bar{q}}(-x,0,t) \), and \( \mathcal{E}^{q}(x,t) = E^{q}(x,0,t) + E^{\bar{q}}(-x,0,t) \).

In the forward limit \( t \to 0 \), the helicity-conserving GPDs reduce to the standard valence quark distributions:
\begin{eqnarray*}
	\mathcal{H}^{u}(x,0) &= u_v(x), \\
	\mathcal{H}^{d}(x,0) &= d_v(x),
\end{eqnarray*}
providing a direct link between GPDs and conventional PDFs.

In contrast, the helicity-flip distributions \( \mathcal{E}^{q}(x,t) \) at \( t = 0 \) do not correspond to known PDFs; instead, they encode novel information about the transverse spin and orbital dynamics of quarks within nucleons. Moreover, to ensure correct falloff behavior at large \( x \), the functional form of \( \mathcal{E}^q(x) \) is modeled with additional suppression near \( x \to 1 \), resulting in a steeper decline than that of \( \mathcal{H}^{q}(x) \)~\cite{Guidal:2004nd,Selyugin:2009ic}.

The normalization of \( \mathcal{E}^{q}(x) \) is fixed by the anomalous magnetic moments:
\begin{equation}
	\kappa_q = \int_{0}^{1} dx\, \mathcal{E}_q(x),
\end{equation}
with
\begin{eqnarray*}
	\kappa_u &= 2\kappa_p + \kappa_n \approx +1.673, \label{eq:9} \\
	\kappa_d &= \kappa_p + 2\kappa_n \approx -2.033, \label{eq:10}
\end{eqnarray*}
where \( \kappa_p = F_2^p(0) = 1.793 \) and \( \kappa_n = F_2^n(0) = -1.913 \).

Additionally, the normalization condition \( \int_{0}^{1} dx\, \mathcal{H}^{q}(x,0) = F_1^{q}(0) \) ensures that for the proton, \( F_1^p(0) = 1 \), while for the neutron, \( F_1^n(0) = 0 \), consistent with electric charge conservation.

To implement the helicity-flip GPDs, we adopt the following parameterizations:
\begin{eqnarray*}
	\mathcal{E}_{u}(x) &= \frac{\kappa_{u}}{N_{u}} (1 - x)^{\eta_{u}} u_v(x), \\
	\mathcal{E}_{d}(x) &= \frac{\kappa_{d}}{N_{d}} (1 - x)^{\eta_{d}} d_v(x), \label{eq:11}
\end{eqnarray*}
where \( \eta_{u,d} \) are falloff parameters and the normalization factors are given by~\cite{Guidal:2004nd}:
\begin{eqnarray*}
	N_u &= \int_0^1 dx\, (1 - x)^{\eta_u} u_v(x), \\
	N_d &= \int_0^1 dx\, (1 - x)^{\eta_d} d_v(x).
\end{eqnarray*}

These definitions provide a consistent and physically motivated framework for constructing \( \mathcal{E}^q(x) \) and ensuring compatibility with both experimental constraints and theoretical expectations at the limits \( x \to 1 \) and \( t \to 0 \).

In the next section, we introduce notable ansatze proposed in previous studies and present the AMA25 parameterization developed in this work.

\begin{figure*}
	\includegraphics[clip,width=0.45\textwidth]{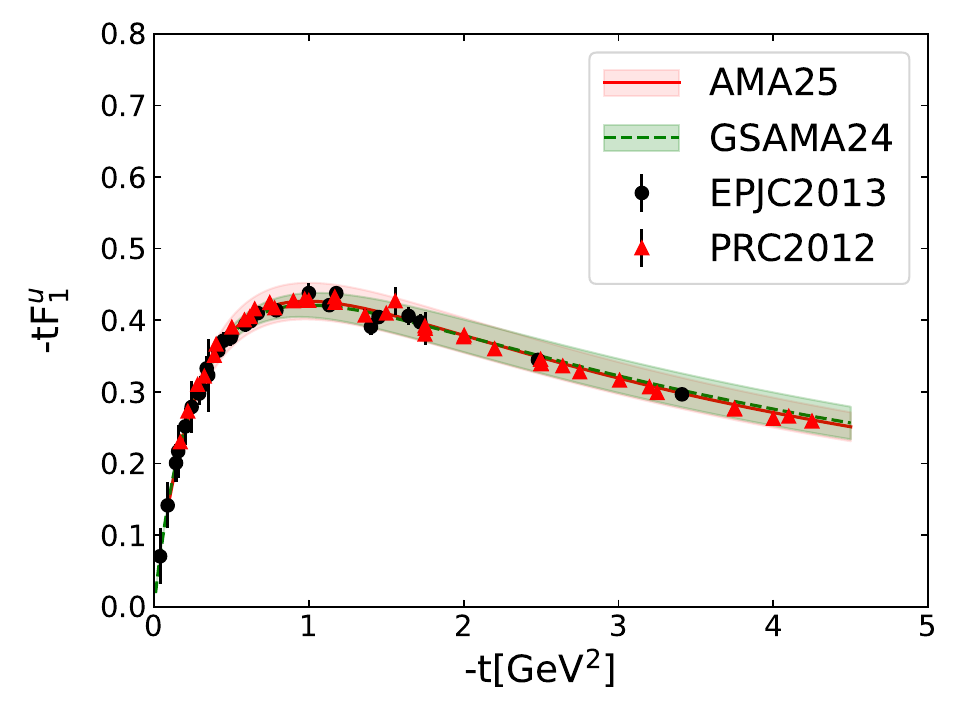 }
	\hspace*{2mm}
	\includegraphics[clip,width=0.45\textwidth]{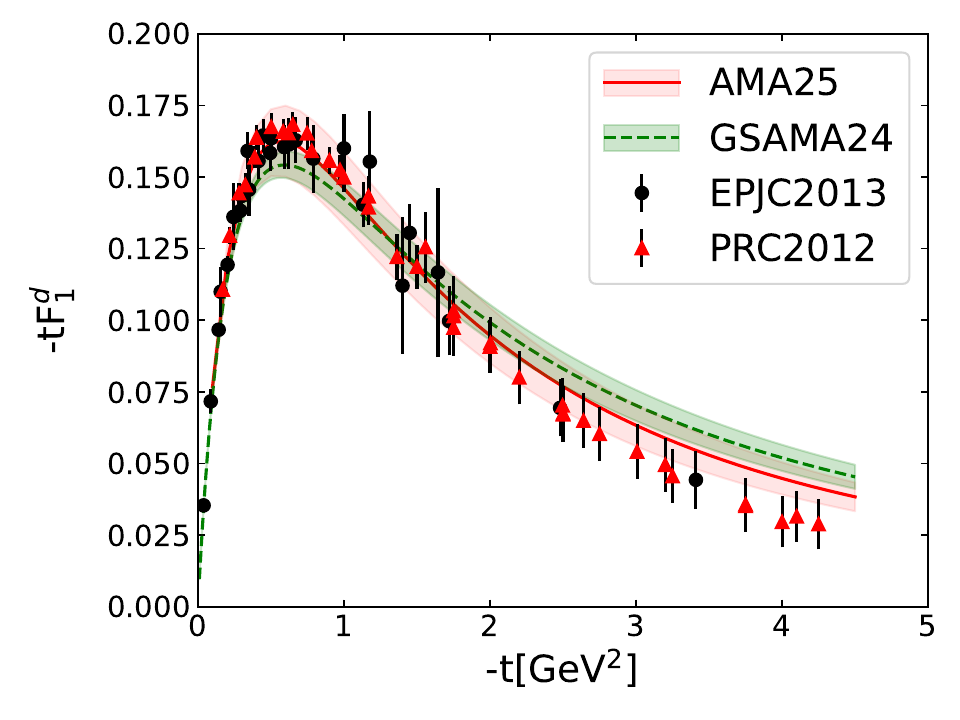 }
	\vspace*{3mm}\\
	\includegraphics[clip,width=0.45\textwidth]{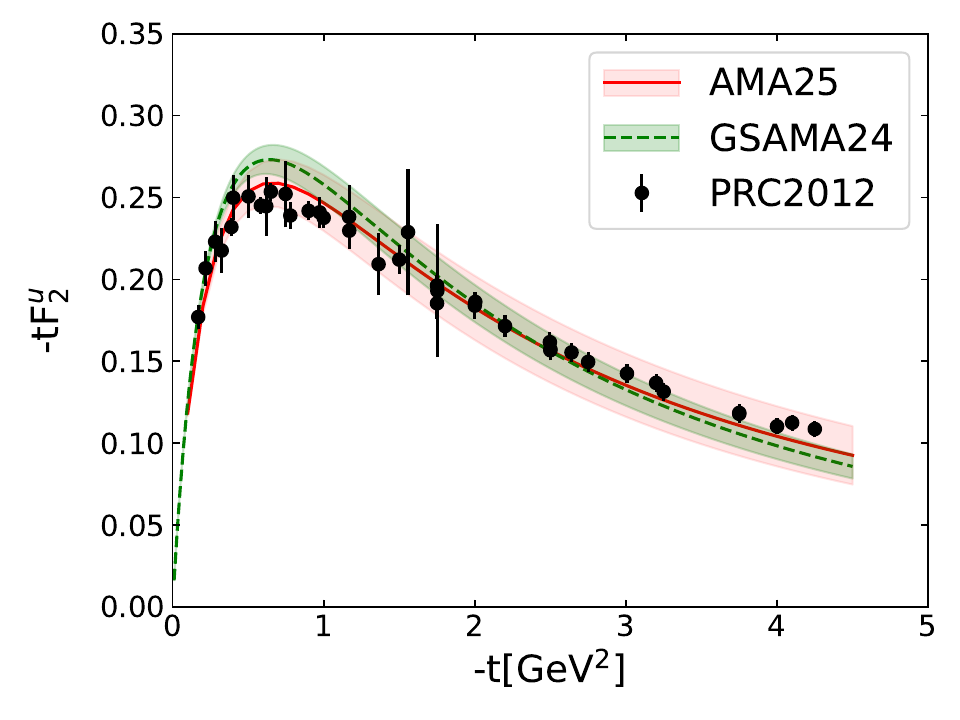 }
	\hspace*{2mm}
	\includegraphics[clip,width=0.45\textwidth]{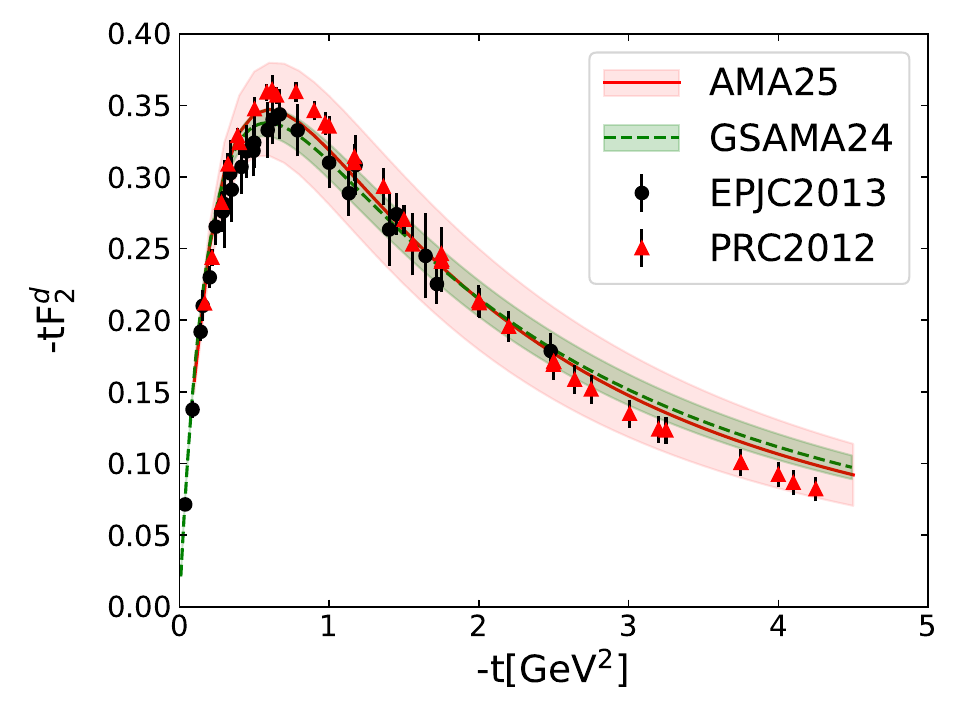 }
	\vspace*{1.5mm}
	
	\caption{\footnotesize The form factors of \(u\) and \(d\) quarks multiplied by \( t \) as a function of \( -t \) using the GSAMA24 ansatz~\cite{Ghasemzadeh:2025qje} and KA08 ~\cite{Khorramian:2008yh} PDF. The points shown are extractions based on experimental data from~\cite{Qattan:2012zf, Diehl:2013xca}. Also GSAMA24 model~\cite{Ghasemzadeh:2025qje} is compared with AMA25 model.}
	\label{fig:tfud1}
\end{figure*}

\section{AMA25 Ansatz and Nucleon Form Factors}\label{sec3}

The AMA25 group has undertaken a systematic effort to refine the description of nucleon form factors by introducing an improved parameterization. The primary objective of this approach is to enhance the accuracy of theoretical predictions while ensuring consistency with experimental observations. The parameterization proposed for this analysis is given by:

\begin{equation}\label{eq:HGS}
	\mathcal{H}^{q}(x,t) = q_{v}(x) \exp \left[ c t (1-x)^g \ln(x) + e x^{b} \ln(1+dt) \right].
\end{equation}

\begin{equation}\label{eq:eGS}
	\varepsilon^{q}(x,t) = \varepsilon_{q}(x) \exp \left[ c t (1-x)^g \ln(x) + e x^{b} \ln(1+dt) \right].
\end{equation}

  Here, $q_v(x)$ represents the valence quark PDF in the forward limit, while $\mathcal{E}_q(x)$ denotes the corresponding helicity-flip distribution. The parameters $c$, $g$, $b$, $e$, and $d$ govern the $t$-dependence of the GPD and are treated as free parameters in the fit.

To validate the effectiveness of the AMA25 ansatz, we applied it to experimental form factor data using a least-squares fitting method. This approach minimizes discrepancies between theoretical predictions and observed values, ensuring a robust comparison. A summary of the dataset utilized in this analysis is provided in Table \ref{tab:tab11}, highlighting its significance as a comprehensive collection of experimental measurements.

To assess the quality of the fit, we computed the reduced chi-squared value, \( \chi^2/{d.o.f} \), which serves as a quantitative measure of agreement between the model and experimental data. The coefficients of Eq.~\ref{eq:HGS} and Eq.~\ref{eq:eGS} were determined through this fitting procedure and are presented in Table \ref{tab:tabresult}. Additionally, the required parameters \( \eta_{u} \) and \( \eta_{d} \) for the AMA25 ansatz are listed in Table \ref{tab:tabresult}.

The values of \( \eta_{u,d} \) exhibit sensitivity to the theoretical framework and the specific methodologies employed in their derivation. Variations in the input dataset can lead to corresponding shifts in these parameters, reflecting the inherent complexities in accurately modeling the internal dynamics of quarks. This underscores the necessity of a careful selection of experimental constraints and theoretical assumptions to ensure the reliability of extracted nucleon form factors.

The AMA25 ansatz provides improved flexibility in capturing the $t$-dependence of GPDs, allowing for a more precise representation of nucleon structure at $\xi = 0$. The refined parameterization enhances computational efficiency while maintaining consistency with fundamental QCD principles. Future investigations will focus on extending this framework to a broader range of kinematic conditions, further improving our understanding of hadronic structure.
Our results underscore the inherent complexity of quark dynamics and interactions within nucleons. While large variations in extracted parameters might initially appear concerning, they are, in fact, indicative of the diverse conditions under which these factors are derived. Such variations reflect the multifaceted nature of particle interactions, emphasizing the sensitivity of nucleon structure to different theoretical frameworks and experimental constraints.

The values of \( \chi^2/{d.o.f} \) for different data sets, summarized in Table \ref{tab:tab3}, provide a direct evaluation of the models performance against a standardized set of experimental measurements. A lower \( \chi^2/{d.o.f} \) value signifies a better fit, facilitating objective comparisons among competing theoretical models.

To further refine our analysis, we decomposed the \( \chi^2/{d.o.f} \) results into distinct components based on separate experimental contributions. Specifically, we categorized the data according to up/down quark distributions and proton/neutron datasets. This segmentation allows for a more granular assessment of how different nucleon substructures influence the overall fit quality. By isolating these contributions, we gain deeper insights into the role of valence quarks and their impact on nucleon form factors.

\begin{figure*}
	\includegraphics[clip,width=0.45\textwidth]{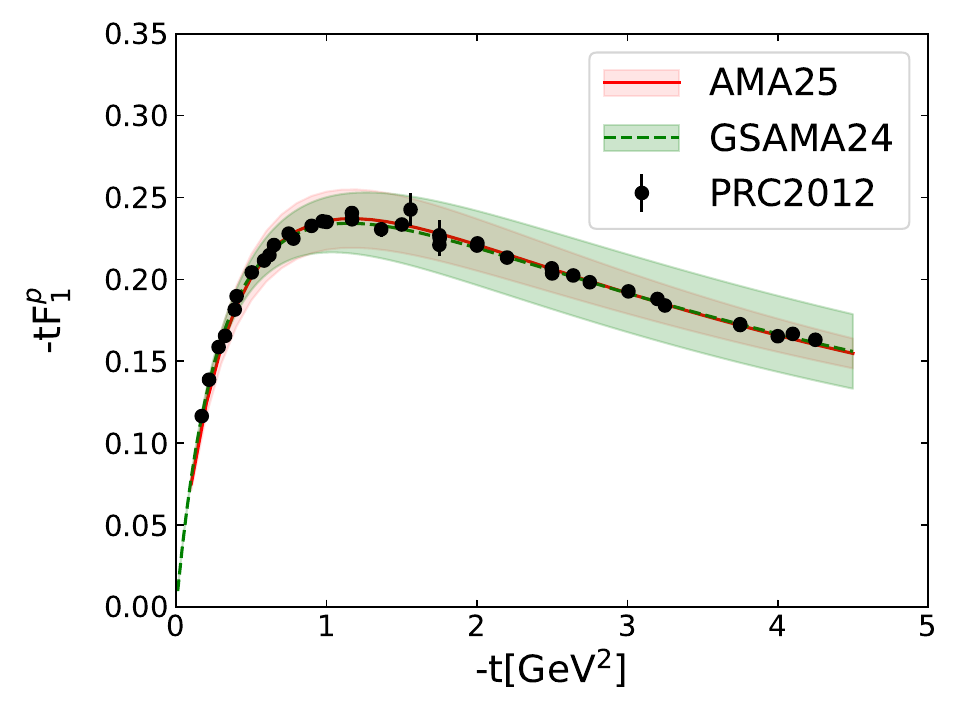 }
	\hspace*{2mm}
	\includegraphics[clip,width=0.45\textwidth]{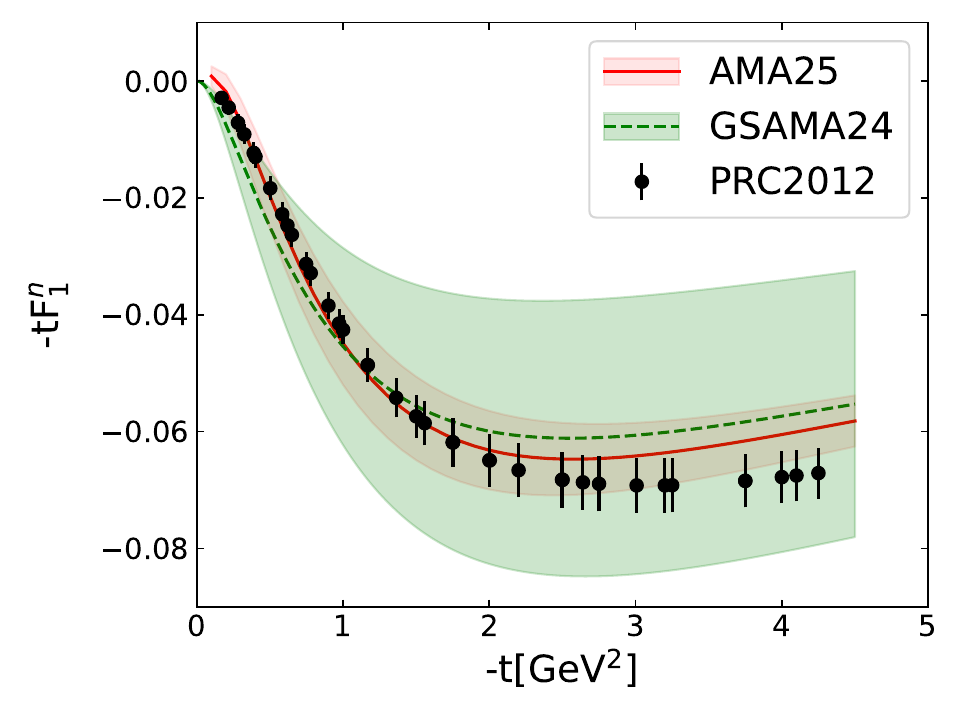 }
	\vspace*{3mm}\\
	\includegraphics[clip,width=0.45\textwidth]{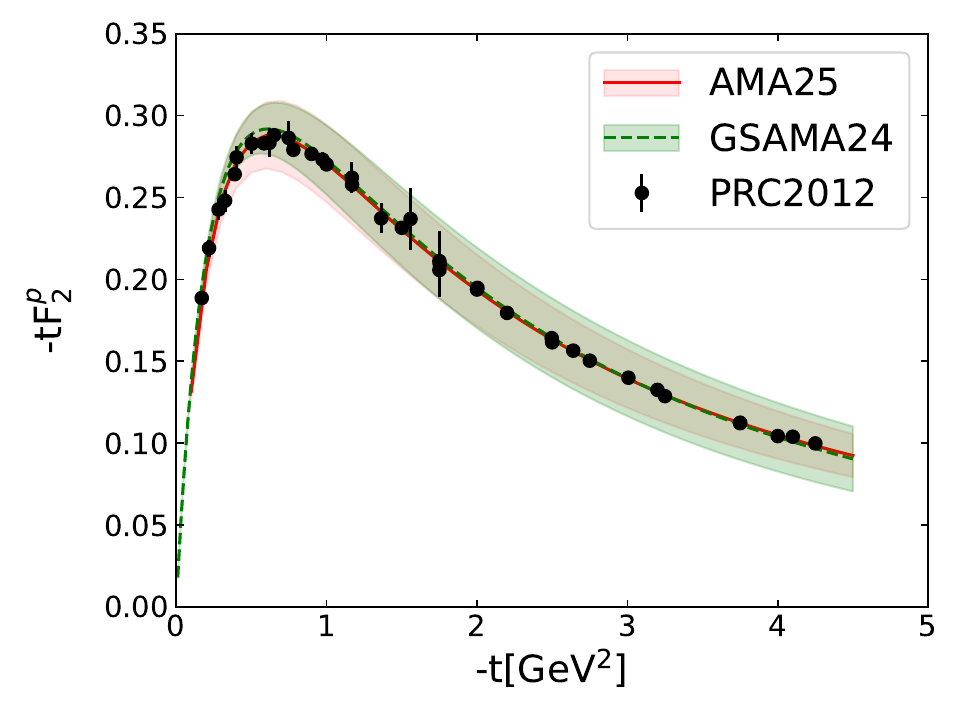 }
	\hspace*{2mm}
	\includegraphics[clip,width=0.45\textwidth]{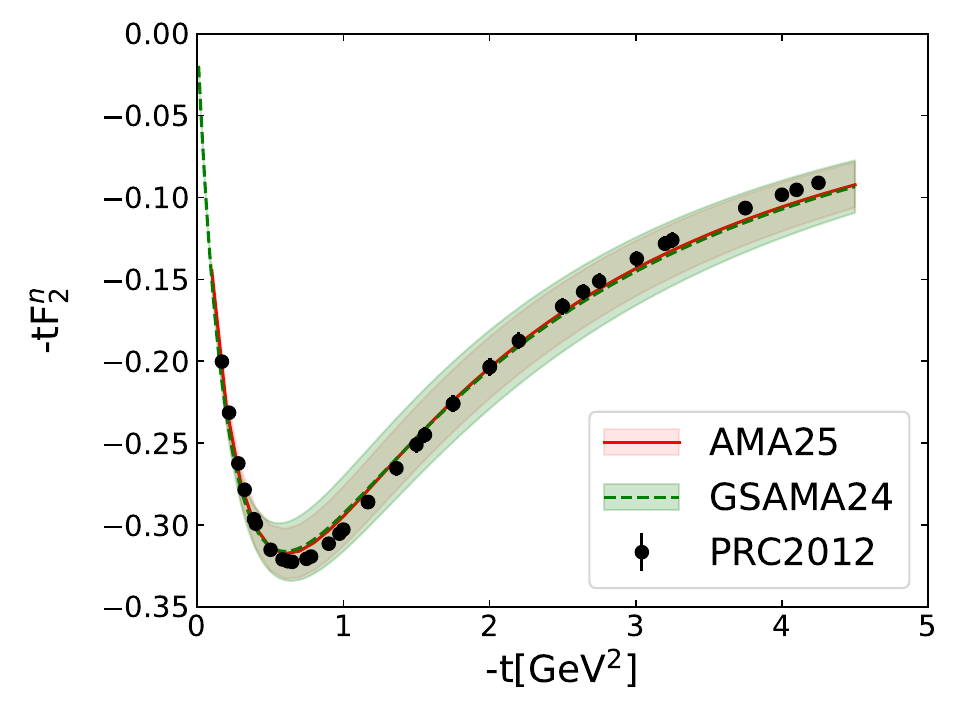 }
	\vspace*{1.5mm}
	\caption{\footnotesize The form factors of proton and neutron  multiplied by $t$ as a function of $-t$ using the GSAMA24 ansatz and KA08~\cite{Khorramian:2008yh}. The points shown are extractions based on experimental data from ~\cite{Qattan:2012zf} (square). Also the results of GSAMA24 model~\cite{Ghasemzadeh:2025qje} is compared with AMA25 model.}
	\label{fig:tfud}
\end{figure*}

\begin{table}
	\begin{center}
		\caption{The data used for analysis.}
		\label{tab:tab11}
\begin{tabular}{lllll}
	\hline
	Dataset & Form Factor & Ref. & t & y \\ \hline
	Dataset 1 & tF1d & \cite{Diehl:2013xca} & 0.039$<$t$<$3.41 & 0.0354%
	$<$y$<$0.164496 \\ 
	Dataset 2 & tF1d & \cite{Qattan:2012zf} & 0.16957$<$t$<$4.25 & 0.103233%
	$<$y$<$0.168285 \\ 
	Dataset 3 & tF1u & \cite{Diehl:2013xca} & 0.039$<$t$<$3.41 & 0.0706%
	$<$y$<$0.438 \\ 
	Dataset 4 & tF1u & \cite{Qattan:2012zf} & 0.16957$<$t$<$4.25 & 0.230098%
	$<$y$<$0.432429 \\ 
	Dataset 5 & tF1p & \cite{Qattan:2012zf} & 0.16957$<$t$<$4.25 & 0.116496%
	$<$y$<$0.242643 \\ 
	Dataset 6 & tF1n & \cite{Qattan:2012zf} & 0.16957$<$t$<$4.25 & -0.069221%
	$<$y$<$-0.002896 \\ 
	Dataset 7 & tF2d &  \cite{Diehl:2013xca} & 0.039$<$t$<$3.41 & 0.071487%
	$<$y$<$0.34371 \\ 
	Dataset 8 & tF2d & \cite{Qattan:2012zf} & 0.16957$<$t$<$4.25 & 0.0823%
	$<$y$<$0.36092 \\ 
	Dataset 9 & tF2u & \cite{Qattan:2012zf} & 0.16957$<$t$<$4.25 & 0.108588%
	$<$y$<$0.253526 \\ 
	Dataset 10 & tF2p & \cite{Qattan:2012zf} & 0.16957$<$t$<$4.25 & 0.099832%
	$<$y$<$0.288022 \\ 
	Dataset 11 & tF2n & \cite{Qattan:2012zf} & 0.16957$<$t$<$4.25 & -0.32251%
	$<$y$<$-0.0910 \\ \hline
\end{tabular}
	\end{center}
\end{table}
\begin{table}
	\begin{center}
		\begin{tabular}{ll}
			\hline
			\multicolumn{2}{l}{AMA25 information} \\ \hline
			a$_{u}$ & 0.4708$\pm $ 0.0018 \\ 
			b$_{u}$ & 3.1332 $\pm $ 0.0035 \\ 
			c$_{u}$ & 3.87 $\pm $ 0.22 \\ 
			d$_{u}$ & 14.36 $\pm $ 0.18 \\ 
			a$_{d}$ & 0.78fixed \\ 
			b$_{d}$ & 0.4708$\pm $0.008 \\ 
			c$_{d}$ & 5 $\pm $ 4 \\ 
			d$_{d}$ & 6.66 $\pm $ 0.3 \\ 
			c & 3.857$\pm $ 0.005 \\ 
			g & 0.138$\pm $ 0.0011 \\ 
			e & 1.3652$\pm $ 0.0014 \\ 
			b & 0.3253$\pm $ 0.0026 \\ 
			d & 7.099 $\pm $ 0.021 \\ 
			$\chi ^{2}/d.o.f$ & 1.5 \\ 
			$\eta _{u}$ & 0.545$\pm $0.005 \\ 
			$\eta _{d}$ & -0.121$\pm $0.012 \\ \hline
		\end{tabular}
		\caption{\footnotesize The numerical result from analysis for GPDs and parton distributions.}
		\label{tab:tabresult}
	\end{center}
\end{table}
\begin{table}[htb]
	\begin{center}
	\caption{The values of $\chi ^{2}/d.o.f$ of the AMA25 ansatz  for data analysis .}
	\label{tab:tab3}
\begin{tabular}{llllll}
	\hline
	\multicolumn{6}{l}{PER-DATASET $\chi ^{2}$ ANALYSIS} \\ \hline
	Dataset   & N pts & Raw $\chi ^{2}$ & Scaled $\chi ^{2}$ & $\chi ^{2}/pt$ & 
	\%total \\ \hline
	Dataset 1 & 27 & 697.9 & 56.45 & 2.091 & 9.5\% \\ 
	Dataset 2 & 39 & 3674.8 & 297.22 & 7.621 & 50.0\% \\ 
	Dataset 3 & 27 & 13.3 & 1.07 & 0.040 & 0.2\% \\ 
	Dataset 4 & 39 & 64.4 & 5.21 & 0.133 & 0.9\% \\ 
	Dataset 5 & 39 & 97.8 & 7.91 & 0.203 & 1.3\% \\ 
	Dataset 6 & 39 & 44.2 & 3.58 & 0.092 & 0.6\% \\ 
	Dataset 7 & 27 & 161.8 & 13.09 & 0.485 & 2.2\% \\ 
	Dataset 8 & 39 & 1541.8 & 124.70 & 3.198 & 21.0\% \\ 
	Dataset 9 & 39 & 901.6 & 72.92 & 1.870 & 12.3\% \\ 
	Dataset 10 & 39 & 25.0 & 2.02 & 0.052 & 0.3\% \\ 
	Dataset 11 & 39 & 132.7 & 10.74 & 0.275 & 1.8\% \\ \hline
	TOTAL & 393 & 7395.3 & 594.90 & 1.514 & 100\% \\ \hline
	\multicolumn{6}{l}{FIT QUALITY ASSESSMENT} \\ \hline
	\multicolumn{6}{l}{Overall reduced $\chi ^{2}=1.514$} \\ 
	\multicolumn{6}{l}{$\chi ^{2}$ per degree of freedom = 1.574} \\ 
	\multicolumn{6}{l}{POOR FITS $(\chi ^{2}/pt>2.0)$ : Dataset 1, Dataset 2,
		Dataset 8} \\ 
	\multicolumn{6}{l}{BEST: Dataset 3 $(\chi ^{2}/pt=0.040)$} \\ 
	\multicolumn{6}{l}{WORST: Dataset 2 \ $(\chi ^{2}/pt=7.621)$} \\ \hline
\end{tabular}
		\end{center}
\end{table}
In this study, we employ the standard Parton Distribution Functions (PDFs) for our analysis, ensuring consistency with established theoretical frameworks. The functional forms of the valence quark distributions are given by:

\begin{equation}\label{eq:xuv}
	xu_v = n_u x^{a_u} (1 - x)^{b_u} \left( 1 + c_u x^{0.5} + d_u x \right),
\end{equation}

\begin{equation}\label{eq:xdv}
	xd_v = n_d x^{a_d} (1 - x)^{b_d} \left( 1 + c_d x^{0.5} + d_d x \right).
\end{equation}

The coefficients for the parton distributions were determined as shown in Table \ref{tab:tabresult}. To ensure proper normalization, we impose the sum rules:

\begin{equation}
	\int_{0}^{1} u_v dx = 2, \quad \int_{0}^{1} d_v dx = 1.
\end{equation}

For parameter optimization, we utilized the iMinuit~\cite{iminuit} package within a Jupyter notebook environment. This approach significantly enhances computational efficiency, allowing for rapid convergence of the fitting procedure. The results obtained from this analysis are summarized in Table \ref{tab:tabresult}, demonstrating the robustness of our parameterization.

The refined parameterization and optimization methodology employed in this study contribute to a more precise representation of nucleon structure. Future investigations will focus on extending this framework to incorporate additional constraints from experimental data, further improving the accuracy of extracted parton distributions.

\begin{figure*}
	\includegraphics[clip,width=0.45\textwidth]{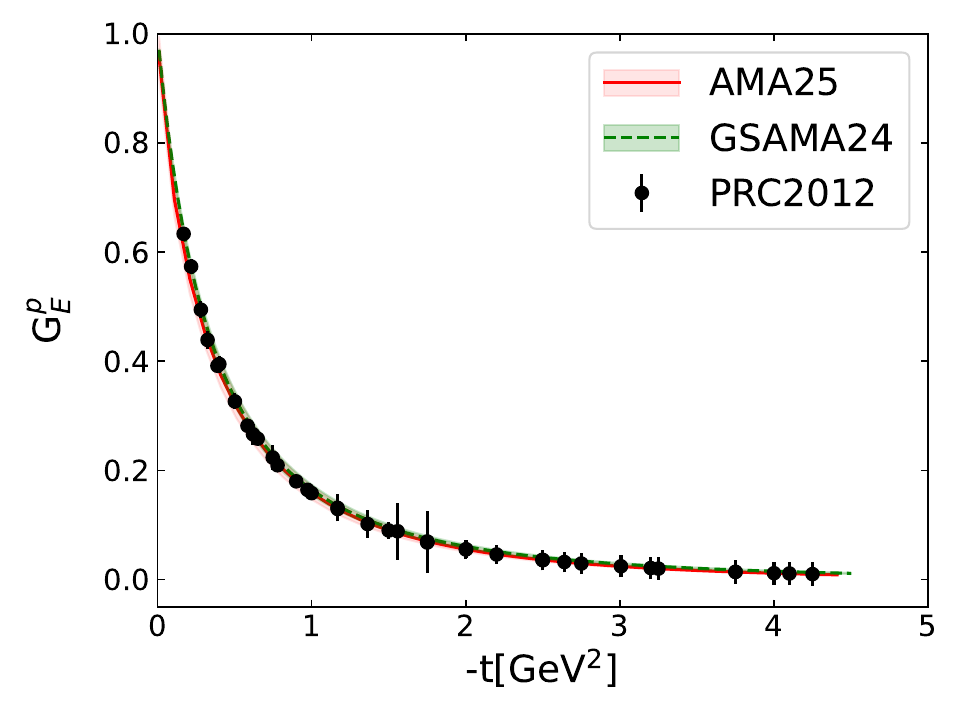}
	\hspace*{2mm}
	\includegraphics[clip,width=0.45\textwidth]{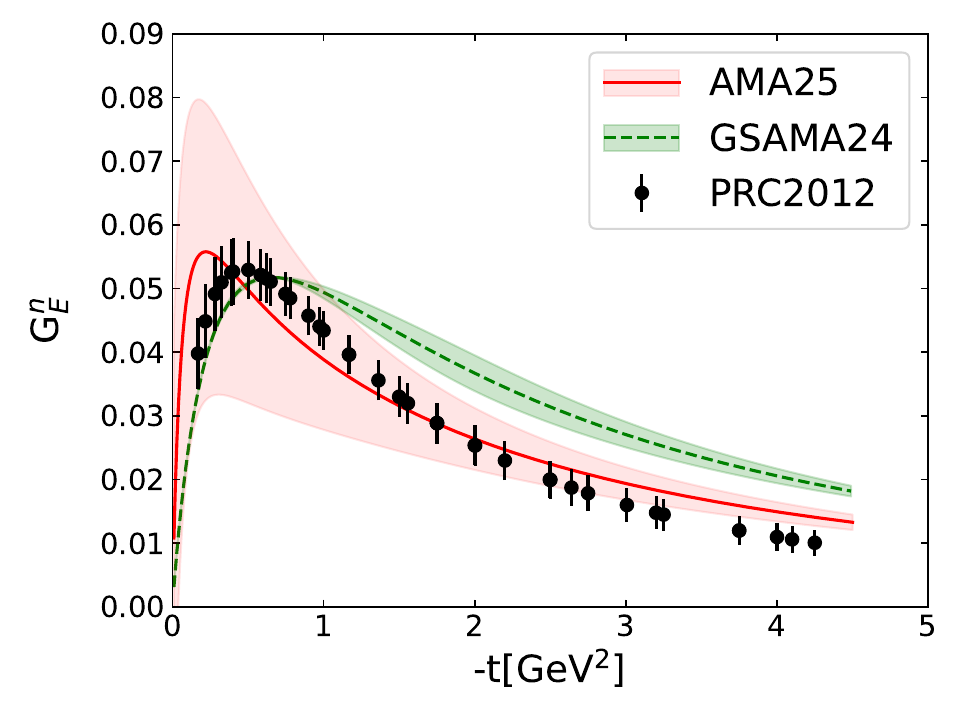 }
	\vspace*{3mm}\\
	\includegraphics[clip,width=0.45\textwidth]{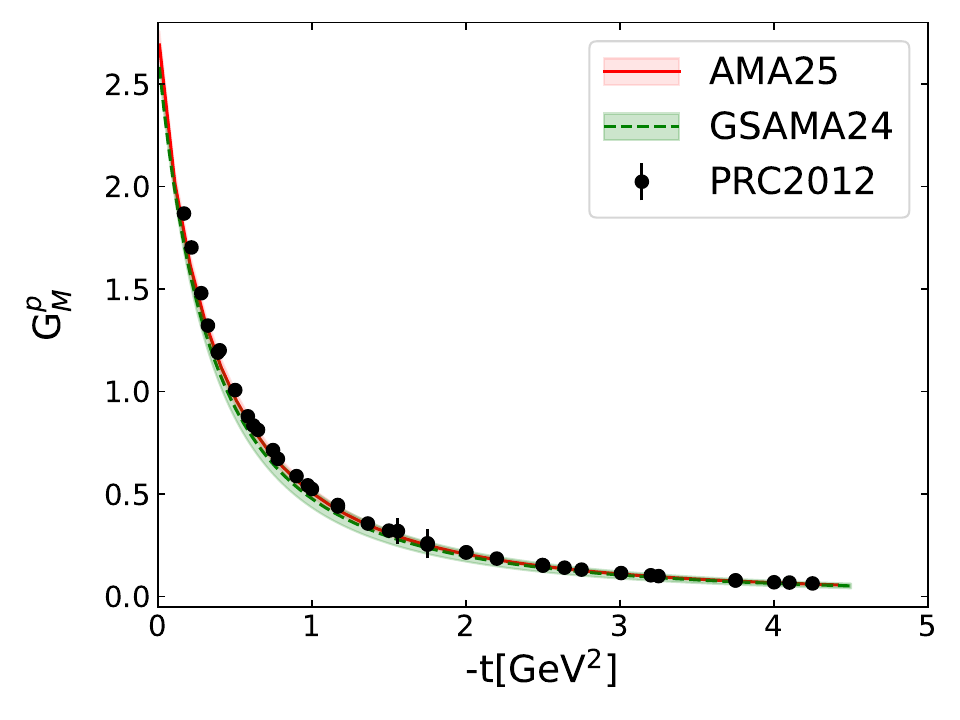 }
	\hspace*{2mm}
	\includegraphics[clip,width=0.45\textwidth]{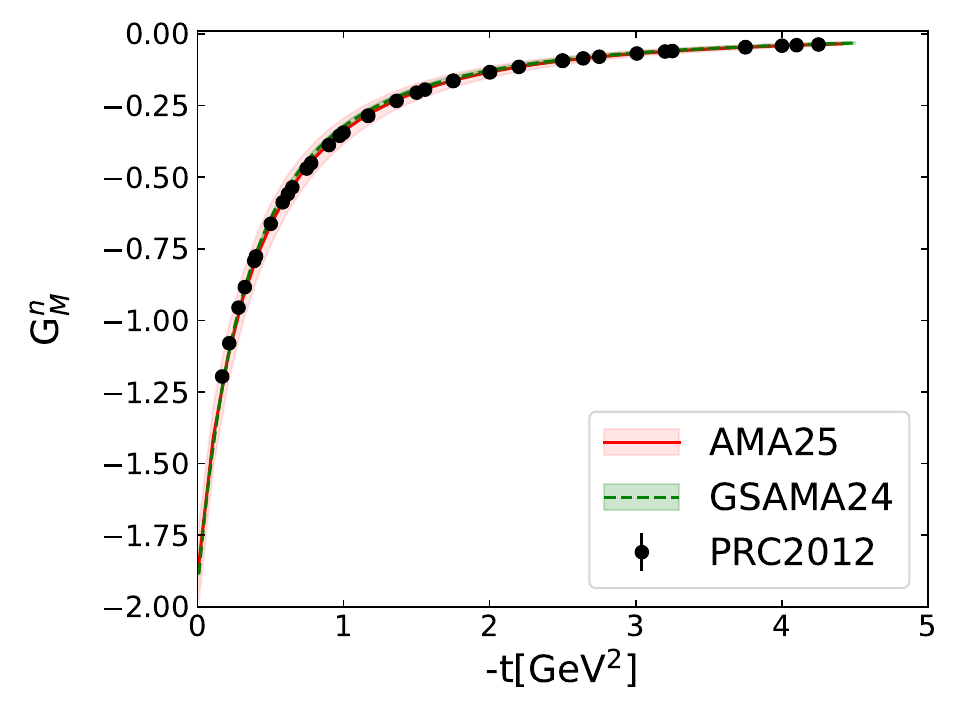 }
	\vspace*{1.5mm}
	\caption{\footnotesize The electric and magnetic form factors \( G_E^{p,n} \) and \( G_M^{p,n} \) as functions of \( t \). The combination of the GSAMA24 model~\cite{Ghasemzadeh:2025qje} is compared with the AMA25. The points shown are extractions based on experimental data from~\cite{Qattan:2012zf} (square).}
	\label{fig:GEMpn}
\end{figure*}

\begin{figure}
	\includegraphics[clip,width=0.45\textwidth]{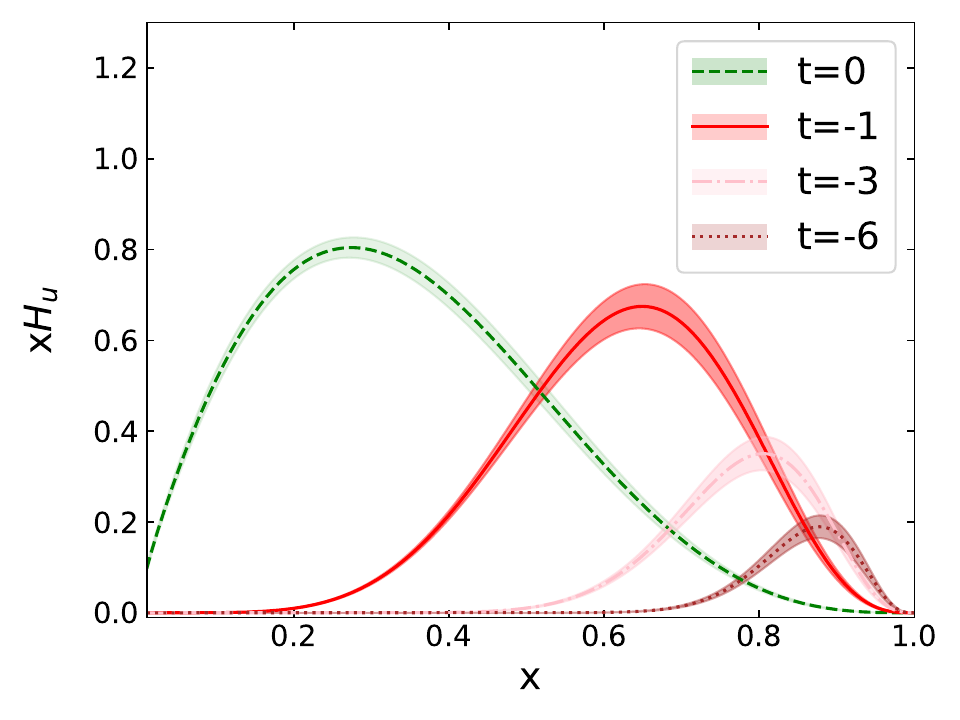 }
	\hspace*{2mm}
	\includegraphics[clip,width=0.45\textwidth]{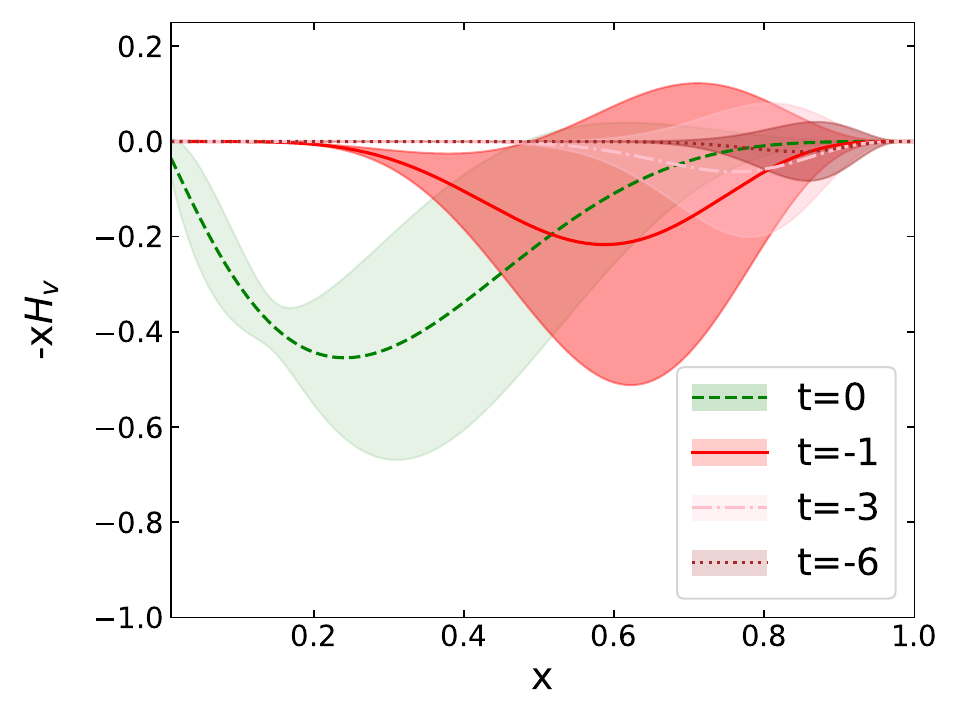 }
	\vspace*{3mm}
	\caption{\footnotesize{ $xH_u$ (quark $u$) and $-xH_v$ (quark $d$) as functions of $t$; see Eq.~\ref{eq:HGS}}}
	\label{fig:Hvu}
\end{figure}
\begin{table*}[htb]
	\begin{center}
		\caption{{\footnotesize The electric radii of the proton and neutron were calculated using KA08~\cite{Khorramian:2008yh} parton distribution functions (PDFs) based on the extended  (GSAMA24)~\cite{Ghasemzadeh:2025qje} models. The data used in this study are obtained from \cite{ParticleDataGroup:2010dbb}.}
			\label{tab:tabR}}
		\vspace*{1.5mm}
		\begin{tabular}{ccc}
			\hline\hline
			PDFs & $r_{E,p}$ & $r^2_{E,n} $   \\   \hline
			Experimental data   & 0.877$\pm$0.009(stat)$\pm$0.011(syst) $fm$ & -0.115$\pm$0.013(stat)$\pm$0.007(syst) $fm^2$ \\
			GSAMA24-KA08   & 0.861893$\pm$0.061841 $fm$ & -0.116988$\pm$0.0025322 $fm^2$ \\
			AMA25   & 0.87866$\pm$0.0023101 $fm$ & -0.11343$\pm$0.013117 $fm^2$ \\
			\hline\hline					
		\end{tabular}
	\end{center}
\end{table*}

With all the necessary components established for calculating nucleon form factors, as outlined in Section \ref{sec:sec2}, we proceed by employing the standard Parton Distribution Functions (PDFs) alongside the parametrized AMA25 ansatz. This framework enables the computation of the Dirac form factor \( F_{1}(t) \) using \( \mathcal{H}(x,t) \) and the Pauli form factor \( F_{2}(t) \) using \( \varepsilon(x,t) \) for the valence quarks \( u \) and \( d \). These calculations allow us to systematically derive the hierarchy of factors associated with protons and neutrons.

By performing these computations, we can illustrate the variations in nucleon form factors as functions of \( t \), specifically highlighting the contributions from \( u \) and \( d \) quarks as well as the proton and neutron. The results of these analyses are presented in Figs.~\ref{fig:tfud1} and \ref{fig:tfud}, demonstrating the dependence of form factors on momentum transfer.

The findings indicate that the AMA25 ansatz exhibits strong compatibility with experimental data, providing improved results compared to the GSAMA24 ansatz across multiple aspects. The enhanced flexibility of AMA25 in capturing the \( t \)-dependence  at zero skewness contributes to a more precise representation of nucleon structure  within this restricted kinematic region. Furthermore, the refined parameterization ensures consistency with fundamental QCD principles while maintaining computational efficiency.

The electric \( G_{E} \) and magnetic \( G_{M} \) form factors play a crucial role in understanding the nucleon's internal structure, as they encapsulate essential information about the spatial charge and current distributions within the nucleon~\cite{Qattan:2012zf, Ernst:1960zza}. These form factors provide insights into the underlying dynamics governing nucleon interactions, reinforcing the predictive power of the AMA25 ansatz.

Future investigations will focus on extending this framework to a broader range of kinematic conditions, further improving the accuracy of extracted nucleon form factors and deepening our understanding of hadronic structure.

\begin{equation}
	G_{E}^{N}(t) = F_{1}(t) + \frac{t}{4M^{2}}F_{2}(t), \;\;\; G_{M}^{N}(t) = F_{1}(t) + F_{2}(t).
	\label{eq:GN}
\end{equation}

The nucleon's form factors are functions of the squared four-momentum transfer \( t \), and in the limit as \( t \to 0 \), they reveal its static properties, such as charge and magnetic moment.

The nonrelativistic interpenetration simplifies the analysis by assuming that nucleon's charge and magnetization are primarily carried by its constituent quarks namely, the up and down quarks. These quarks are thought to exhibit similar spatial distributions, leading to comparable contributions to the form factors.

  Investigating the individual contributions of up and down quarks to nucleon form factors provides insights into the underlying principles of Quantum Chromodynamics (QCD). Critically, as noted below, these contributions incorporate both quark and antiquark components, representing the charge-weighted difference between their distributions \cite{Perdrisat:2006hj}, a distinction that is pivotal for accurately interpreting nucleon form factors.

Ongoing research and experimentation continue to shed light on the complexities of nucleon structure, revealing the intricate forces at play within atomic nucleons. The insights gained from these investigations not only refine theoretical models but also provide avenues for technological innovations rooted in quantum mechanics and nuclear physics \cite{Beck:2001yx,Miller:1990iz}

  Within the framework of the Standard Model, the form factors $G_{E,M}^{u}$ and
$G_{E,M}^{d}$ encapsulate the contributions of up and down quarks to the
proton's and neutron's electric and magnetic properties. These form factors are
fundamental in quantifying the intrinsic electromagnetic characteristics of
these subatomic particles. Using the quark model and isospin symmetry, the
electromagnetic form factors of the nucleons can be expressed directly in terms
of these quark form factors. Considering the quark composition of the proton
($uud$) and the neutron ($udd$), these relations are:
\begin{eqnarray}
	G^p_{E,M} &=& \frac{2}{3}G^u_{E,M} - \frac{1}{3}G^d_{E,M}, \nonumber \\
	G^n_{E,M} &=& \frac{2}{3}G^d_{E,M} - \frac{1}{3}G^u_{E,M}.
\end{eqnarray}
Similarly, the Dirac and Pauli form factors, $F_1$ and $F_2$, obey analogous
relations, reflecting the underlying symmetry in quark contributions to nucleon
structure.

The magnetic moments of the constituent quarks, denoted as $\mu_{u}$ and $\mu_{d}$, can be inferred from the behavior of the magnetic form factors in the limit as the momentum transfer approaches zero. Specifically, they are given by:

\begin{eqnarray}
	\mu_{u} &=& 2\mu_{p} + \mu_{n} = 3.67\mu_{N}, \\ 
	\mu_{d} &=& \mu_{p} + 2\mu_{n} = -1.03\mu_{N}.
\end{eqnarray}

Here, $\mu_{N}$ represents the nuclear magneton, a fundamental constant that serves as the standard unit for nucleon magnetic moments. This distinction is essential as it captures the charge-weighted difference in their contributions to nucleon form factors.

This refined interpretation is essential for accurately contextualizing both experimental data and theoretical frameworks.  Figure \ref{fig:GEMpn} presents the proton and neutron electric ($G_{E}$) and magnetic ($G_{M}$) form factors obtained from the AMA25 and GSAMA24 ansatz models, providing insight into the structure of the nucleon.

 Furthermore, in Figure~\ref{fig:Hvu}, we present the extracted distributions $xH_u$ for quark $u$ and $-xH_v$ for quark $d$ as functions of $t$, as described in Eq.~\ref{eq:HGS}, for $t = 0, -1, -3, -6$. These results provide valuable insights into the $t$-dependence of the valence quark GPDs, reinforcing the predictive power of our model.

The form factors are plotted as functions of the negative squared momentum transfer $t$, illustrating their evolution across interaction energy scales. The intricate relationship between quark distributions and the resulting electromagnetic form factors remains a central focus of research. By leveraging diverse models and distribution functions, we seek to deepen our understanding of nucleon substructure and the fundamental forces governing its behavior.

Our findings indicate that the AMA25 model shows improved consistency with experimental data compared to the GSAMA24 model. As illustrated in Figures \ref{fig:tfud1} and \ref{fig:tfud}, this improvement is evident in the form factors of the $u$ and $d$ quarks, as well as in the proton and neutron form factors scaled by $t$.

  Furthermore, Figure 3, which compares electric and magnetic form factors to PRC 2012 experimental data \cite{Qattan:2012zf}, reinforces that the AMA25 model yields the most reliable fit.

The computation of $G^E_n$ is inherently dependent on assumptions regarding nucleon structure and quark interaction dynamics. Accordingly, variations in model parameters or approximations can introduce deviations in predicted values. While agreement for $G^E_n$ remains robust at low $x$, discrepancies at higher $x$ likely stem from the reliance on proton parton distribution functions (PDFs), which exert a pronounced influence on the neutrons $G^E_n$.

 To further refine our analysis and improve agreement with experimental data for both protons and neutrons, we plan to conduct a joint QCD study incorporating both ansatz models and generalized parton distributions (GPDs). This approach aims to enhance theoretical precision while maintaining consistency with empirical observations.

Throughout the paper, we adopt the spacelike convention $t \equiv -Q^2 < 0$. In terms of the Dirac and Pauli form factors, the Sachs form factors are given by~\cite{Masjuan:2012sk}

\begin{equation}
	G_E^{p,n}(t)=F_1^{p,n}(t)-\frac{t}{4M_N^2}\,F_2^{p,n}(t),
	\label{eq:sachs1}
\end{equation}

\begin{equation}
	G_M^{p,n}(t)=F_1^{p,n}(t)+F_2^{p,n}(t).
	\label{eq:sachs2}
\end{equation}

A convenient large-$N_c$ representation uses products of monopoles~\cite{Masjuan:2012sk}:
\begin{eqnarray}
		F_1^{(0)}(t) &= \frac{1 - c_0\, t/m_{\omega''}^2}
		{(1 - t/m_\omega^2)(1 - t/m_{\omega'}^2)(1 - t/m_{\omega''}^2)},\nonumber\\
		F_1^{(1)}(t) &= \frac{1 - c_1\, t/m_{\rho''}^2}
		{(1 - t/m_\rho^2)(1 - t/m_{\rho'}^2)(1 - t/m_{\rho''}^2)},\nonumber\\
		F_2^{(0)}(t) &= \frac{1}
		{(1 - t/m_\omega^2)(1 - t/m_{\omega'}^2)(1 - t/m_{\omega''}^2)},\nonumber\\
		F_2^{(1)}(t) &= \frac{1}
		{(1 - t/m_\rho^2)(1 - t/m_{\rho'}^2)(1 - t/m_{\rho''}^2)},
\end{eqnarray}\label{eq:Fmodel}

The four quantities are isoscalar and isovector vector-meson-dominance (VMD) form factors.

They represent the decomposition of a hadronic current into two independent Lorentz structures (1 and 2), each split into:
\begin{itemize}
	\item $F^{(0)}$ (isoscalar): the component that couples to isoscalar vector mesons (e.g., the $\omega$ meson and its excited states $\omega'$, $\omega''$); 
	\item $F^{(1)}$ (Isovector): the component that couples to isovector vector mesons (e.g., the $\rho$ meson and its excited states $\rho'$, $\rho''$).)
\end{itemize}

The constants $c_0$ and $c_1$ are fixed by vector-meson-nucleon couplings~\cite{Masjuan:2014sua}:94
3

3

0
\begin{eqnarray}
		\frac{g_{\omega NN} f_{\omega\gamma}}{m_\omega^2}
		&= \frac{1}{2}\,\frac{1 - c_0\, m_\omega^2/m_{\omega''}^2}
		{(1 - m_\omega^2/m_{\omega'}^2)(1 - m_\omega^2/m_{\omega''}^2)},\nonumber\\
		\frac{g_{\rho NN} f_{\rho\gamma}}{m_\rho^2}
		&= \frac{1}{2}\,\frac{1 - c_1\, m_\rho^2/m_{\rho''}^2}
		{(1 - m_\rho^2/m_{\rho'}^2)(1 - m_\rho^2/m_{\rho''}^2)}.
\end{eqnarray}	\label{eq:c0c1}

\section{Dirac Mean Squared Radii}\label{sec4}

In the framework of parton distribution functions (PDFs), the Dirac mean-squared radii serve as fundamental parameters quantifying the spatial extent of a proton or neutron based on the distribution of its constituent partons. These radii are extracted through a rigorous analysis of form factors and provide direct insights into nucleon structure at low-energy scales. Notably, within the PDF formalism, the Dirac mean-squared radius is intrinsically linked to the isoscalar electric charge form factor, offering a precise characterization of the nucleon's charge distribution.

A detailed analysis of the Dirac mean-squared radii is essential for isolating distinct structural contributions to nucleon composition, including the quark core and surrounding meson cloud. This decomposition is particularly relevant for understanding nucleon dynamics across different energy scales, as the interplay between short-range quark interactions and long-range pion fields remains a central theme in hadronic physics. The Dirac radii thus serve as indispensable observables for probing the intrinsic spatial structure of nucleons, with implications ranging from low-energy nuclear phenomena to high-energy QCD.

Furthermore, the computed form factors and their low-\( t \) extrapolations were evaluated against recent lattice QCD results~\cite{Alexandrou:2020okk} and dispersive analyses~\cite{Hoferichter:2016duk}, ensuring that the extracted radii remain consistent with well-established theoretical benchmarks. This validation underscores the relevance of the AMA25 ansatz within the broader context of QCD phenomenology.

Our results firmly establish the Dirac mean-squared radii as robust descriptors of nucleon size.  Table~\ref{tab:tabR} presents the extracted values of the meson-nucleon coupling constants, which support the reliability of our parametrization in the small $-t$ region. Notably, the successful extraction of these quantities using the AMA25 framework highlights its predictive strength and confirms the value of employing flexible, physically motivated ansatze in nucleon structure studies.

 Moreover, our methodology provides a systematic approach for disentangling the quark-core contributions from meson-cloud effects, with the inclusion of the vector-meson nucleon contribution partially accounting for the latter. However, we acknowledge that a complete separation would require a full treatment of sea quarks and gluons, which is beyond the scope of this work. This comprehensive treatment strengthens the reliability of our results and confirms the efficacy of our approach in probing nucleon structure at varying momentum scales.

Given the observed consistency between our computed Dirac mean-squared radii and established experimental results, our findings underscore the necessity of integrating such refined analyses into future studies of nucleon substructure. These results not only validate our computational framework but also pave the way for further advancements in nucleon tomography, particularly within the context of generalized parton distributions (GPDs) and QCD phenomenology.

\section{Conclusion}\label{sec:conclusion}

In this study, we presented a novel analysis employing new ansatzes and parton distribution functions (PDFs) to systematically investigate the Pauli, Dirac, and electromagnetic form factors of nucleons. Specifically, we utilized the GSAMA24 ansatz~\cite{Ghasemzadeh:2025qje} and the AMA25 ansatz, incorporating refined parameterizations to enhance accuracy in modeling nucleon substructure.
The free parameters of the AMA25 ansatz and PDF were determined through rigorous analysis of experimental data for $tF_1^u$, $tF_1^d$, $tF_1^p$, $tF_1^n$, $tF_2^u$, $tF_2^d$, $tF_2^p$, and $tF_2^n$. The Dirac and Pauli form factors for up and down quarks specifically, $F_1^u$, $F_1^d$, $F_2^u$, and $F_2^d$ were examined as functions of $-t$, as shown in Fig.~\ref{fig:tfud1}. Our findings demonstrate that the AMA25 ansatz, particularly in conjunction with the PDF obtained through our analysis, shows improved agreement with experimental data from~\cite{Qattan:2012zf,Diehl:2013xca} compared to the GSAMA24 ansatz~\cite{Ghasemzadeh:2025qje}.

Next, we employed the PDF extracted from our analysis using the AMA25 ansatz, as depicted in Fig.~\ref{fig:tfud}. Furthermore, Fig.~\ref{fig:GEMpn} presents proton charge and magnetization densities derived from the AMA25 ansatz with PDFs obtained through our study. These densities were computed using multiple parametrizations and compared with results from previous works, demonstrating strong consistency. To calculate the transverse charge and magnetization density, we utilized the $G_E$ and $G_M$ equations, incorporating experimental data from~\cite{Qattan:2012zf}.

The AMA25 ansatz was specifically developed to refine the theoretical depiction of hadronic internal structure, integrating advanced parameterization techniques to capture both short-range and long-range contributions to nucleon form factors. By systematically applying the AMA25 ansatz and optimizing it against experimental data, we derived crucial insights into the spatial and momentum distributions of quarks and gluons.

Our approach not only validates theoretical predictions but also provides a robust framework for future investigations into nucleon tomography and high-energy QCD phenomenology. The calibration of free parameters via this methodology enhances precision in describing nucleon structure and offers a foundational basis for extending the analysis to generalized parton distributions (GPDs) and deep inelastic scattering studies.

Overall, the results presented in this study reinforce the efficacy of the AMA25 ansatz and establish its superiority in reproducing experimentally observed nucleon form factors. The findings contribute significantly to the ongoing discourse in hadronic physics, offering a refined perspective on nucleon dynamics and advancing the theoretical framework necessary for future experimental validations.

\section*{Data Availability Statement} All data and source codes used in this work are publicly available at: 
\url{https://github.com/atashbart/GPDAtahbar}
	%
	%
\section*{Acknowledgments}
F.~A. acknowledges the Farhangian University for the provided support to conduct this research. S.A.T  grateful to the School of Particles and Accelerators, Institute for Research in Fundamental Sciences (IPM). 
%
%
%

	%
	%
\end{document}